\documentclass[%
reprint,
aps,
prl,
amsmath,
twocolumn,
amssymb,
floatfixng,
showpacs,
superscriptaddress,
footinbib,
multicol
]{revtex4-2}
\usepackage{graphicx}
\usepackage{epstopdf}
\usepackage{comment}
\usepackage{dcolumn}
\usepackage{bm}
\usepackage{color}
\usepackage{xcolor}
\usepackage{amsmath}
\usepackage{braket}
\usepackage{amsfonts}
\usepackage{soul}
\usepackage{url}
\usepackage{array,tabularx}
\usepackage{graphics}
\usepackage{graphicx}
\usepackage{subcaption}
\usepackage{times}

\usepackage{caption}
\DeclareCaptionJustification{fulljust}{\leftskip=0pt\relax\rightskip=0pt\relax\parfillskip=0pt plus 1fil\relax}
\usepackage[export]{adjustbox}
\usepackage[colorlinks=true,linktoc=page,linkcolor=red,citecolor=blue]{hyperref}

\usepackage{xurl}      %
\usepackage{hyperref}  %

\definecolor{OliveGreen}{rgb}{0,0.6,0}

\begin{document}

\title{
Acoustic Scattered Field Calculation for a Moving Source Using Time-Dependent Physical Optics}

\author{Asaf Kor}\thanks{These authors contributed equally to this work.}
\affiliation{School of Mechanical Engineering, Tel Aviv University, Tel Aviv 69978, Israel}

\author{Gal Ben Yelid}\thanks{These authors contributed equally to this work.}
\affiliation{School of Mechanical Engineering, Tel Aviv University, Tel Aviv 69978, Israel}

\author{Amir Boag}\email{boag@tauex.tau.ac.il}
\affiliation{School of Electrical and Computer Engineering, Tel Aviv University, Tel Aviv 69978, Israel}

\author{Lea Beilkin-Sirota}\email{leabeilkin@tauex.tau.ac.il}
\affiliation{School of Mechanical Engineering, Tel Aviv University, Tel Aviv 69978, Israel}

\begin{abstract}
We present a time-dependent physical optics (TDPO) formulation for computing the acoustic scattered field of a stationary object in response to a moving source in three-dimensional space. Due to the motion of the source, which may be fast, the conventional frequency-domain PO method is not employed. Instead, a time-domain variant that explicitly accounts for the motion is formulated. The study focuses on the near-field regime, where source motion induces a complex wave field beyond a simple Doppler effect. The proposed algorithm accommodates arbitrary source trajectories and arbitrary signal bandwidth, as well as high source velocities relative to the speed of sound. Near and beyond the shadow boundary, where PO is least accurate, the formulation is supplemented by a Fock correction of the surface field, extended here to a moving source with a harmonic signal. The method matches or exceeds the accuracy of finite element simulations at a computational cost lower by orders of magnitude.
\end{abstract}

\maketitle

\section{Introduction} \label{sec:intro}

Wave scattering plays a central role in both acoustics and
electromagnetics and underpins a broad range of sensing and characterization
applications. In acoustics, scattered fields are used for environmental
mapping and navigation in air, such as in robotic systems
\cite{jansen2024semantic,laurijssen2024hiris}, as well as for object detection,
underwater navigation, bathymetric mapping, and environmental monitoring in
marine environments
\cite{hodges2011underwater,Ziomek2023,aubard2025sonar,Sun2021,Miller2021,
testa2018prediction,sathish2022investigation}. The analysis and simulation of
such sensing scenarios therefore call for a fast computation of these fields,
with an accuracy sufficient for the application at hand, since both the objects
and the propagation distances involved are typically large compared with the
wavelength.
In this work, we consider the field scattered by a rigid
object, with particular emphasis on the formulation for a moving source.

From an analytical viewpoint, scattering is classically formulated as a
boundary-value problem for the Helmholtz equation, where boundary conditions
are imposed on the scatterer surface (e.g., sound-hard or sound-soft
conditions), together with a radiation condition at infinity to ensure
outgoing waves. A central role is played by Green-function representations
and the Kirchhoff--Helmholtz integral, from which surface integral formulations
of the scattered field can be derived \cite{MorseIngard1968,ColtonKress2019}.
For canonical geometries, exact solutions can further be obtained by separation
of variables and multipole expansions \cite{bowman1970scattering,Martin2010}. In one dimension, explicit time-domain solutions are available even with damped boundaries \cite{sirota2013extended,sirota2014complete}.

When an analytical solution is not available, as is generally the case for
realistic scattering problems, numerical methods must be employed. For
stationary configurations, this is typically carried out in the frequency
domain under time-harmonic excitation. The finite element method (FEM), for example, discretizes the governing
equations throughout the computational domain and is well suited to complex
geometries and inhomogeneous media
\cite{Zienkiewicz2005,Ihlenburg1998}. Its volumetric nature, however, can lead
to a large number of degrees of freedom at high frequencies.

The boundary element method (BEM)
\cite{marburg2008computational} instead reformulates the problem as an
integral equation posed on the scatterer surface. For exterior scattering in
a homogeneous medium, this reduces the dimensionality of the discretization
and inherently accounts for radiation into the unbounded domain. These
advantages come at the cost of generally dense system matrices, arising from
the nonlocal coupling between surface elements,
although 
considerable computational acceleration can be achieved
through fast algorithms
\cite{chernokozhin2016fast,chernokozhin2025numerical}.

For transient problems, time-domain formulations of FEM and BEM can be used,
while the finite-difference time-domain (FDTD) method directly discretizes the
wave equation in space and time
\cite{Taflove2005,rumpf2014finite,vasmel2013immersive}. FDTD uses a structured
grid and local time updates, avoiding the solution of a global matrix system.
However, similarly to FEM, it requires discretization of the entire propagation domain, and grid
anisotropy introduces a direction-dependent wave speed, leading to numerical
dispersion \cite{cheng2021investigation}. Its Cartesian grid also makes complex
geometries less natural to represent and may require conformal or
immersed-boundary techniques \cite{Farjadpour2006}. 

For acoustically large objects, the computational cost of full-wave methods
can become prohibitive, motivating the use of the physical optics (PO)
approximation, also known as the Kirchhoff approximation
\cite{MorseIngard1968,ColtonKress2019,Balanis2012,ufimtsev2014fundamentals}.
In PO, each illuminated surface point is treated according to its local tangent
plane; for an acoustically rigid scatterer, this amounts to taking the total
surface pressure as twice the incident pressure, while on the shadowed part of
the object it is set to zero. Interactions between distinct surface points are
neglected. 

These assumptions are most appropriate when the object dimensions
and local radii of curvature are large compared with the wavelength, and
multiple scattering is weak. 
For convex objects, the resulting abrupt
transition between the lit and shadowed parts can be smoothed by the Fock
theory \cite{Fock1965}, which has been incorporated into ray-based diffraction
methods \cite{Pathak1980,DuTakada2023} and is applied here within PO. The PO surface integral is computationally
efficient, with further acceleration possible through fast implementations
\cite{Boag2004fast}.

The treatment of moving sources introduces additional complexity. Whereas
stationary scattering problems are typically solved in the frequency domain,
arbitrary source motion requires a time-domain formulation. In
time-domain FEM, a moving source may be represented on a fixed mesh by
distributing its forcing, at each time step, among the nodes surrounding its
current position \cite{Manduzio2021}; moving-mesh or remeshing approaches may
also be employed \cite{Donea2004ALE}. Likewise, in FDTD, the source must be
mapped onto the fixed grid at each time step, by interpolation or by
distribution over neighboring grid points
\cite{TsuchiyaKanamori2021,dragna2014modeling}. In time-domain BEM, which is
formulated with retarded potentials \cite{takahashi2022enhancement}, the
incident field on the boundary depends on the source position at the retarded
emission time, defined implicitly by the source trajectory
\cite{AntesMeise1991}.

These considerations motivate a direct time-domain formulation of PO for
moving-source scattering. Time-domain PO has been formulated for stationary
configurations \cite{sun1994time}, and frequency-domain PO has been applied to
harmonic sources moving slowly at constant speed, by evaluating the surface
integral at the Doppler-shifted frequencies \cite{pignier2015kirchhoff}. Here,
PO is formulated directly in the time domain, as time-dependent PO (TDPO), for arbitrary source
trajectories and transient signals, and the Fock correction is extended to a
moving source with a harmonic signal (Sec.~\ref{sec:theory}). 

The numerical
evaluation is described in Sec.~\ref{sec:numerical}. In Sec.~\ref{sec:sim},
the method is validated against the exact solution for a shape that admits
one and compared with FE, for a stationary and a moving source, and then
compared with a transient FE simulation for a shape for which no exact solution
is available.

\section{Theory}  \label{sec:theory}

\begin{figure}[!tb]
    \centering   
          \includegraphics[width = 0.9\columnwidth]{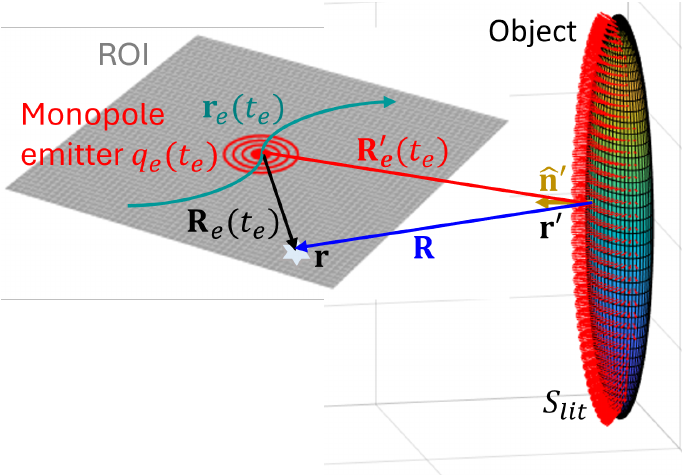}  
  \caption{\textbf{General scheme.} A moving point source acoustically illuminates a stationary rigid object. The time-dependent quantities are written in terms of the emission time $t_e$. Normal unit vectors are shown on $S_{lit}$.}
    \label{fig:scat_field_scheme}
\end{figure}

We consider a monopole source $q_e(t_e)$ illuminating a perfectly rigid and stationary object bounded by the surface $S$, as illustrated in Fig.~\ref{fig:scat_field_scheme}. 
The propagation takes place in a homogeneous, isotropic, and lossless acoustic medium characterized by constant mass density $\rho_0$ and sound speed $c$ (e.g., air or water).
In the general case, the source follows a time-dependent trajectory $\mathbf{r}_e(t_e)$, defining the vectors $\mathbf{R}_e(t_e) = \mathbf{r} - \mathbf{r}_e(t_e)$ to an observation point $\mathbf{r}$ in the region of interest (ROI) and $\mathbf{R}_e'(t_e) = \mathbf{r}' - \mathbf{r}_e(t_e)$, with the respective magnitudes $R_e=|\mathbf{R}_e|$ and $R'_e=|\mathbf{R}'_e|$, to a surface point $\mathbf{r}' \in S$. Here, $t_e$ denotes the emission time, often referred to as the retarded time.
The illuminated portion of the surface, $S_{\mathrm{lit}}$, is defined by $\mathbf{R}_e'(t_e) \cdot \widehat{\mathbf{n}}' < 0$, where $\widehat{\mathbf{n}}'$ is the outward unit normal at $\mathbf{r}'$.  
The distance between a surface point and the observation point, given by $R=|\mathbf{R}|$, where $\mathbf{R} = \mathbf{r} - \mathbf{r}'$, is time-invariant since the object is stationary.  

In the absence of the object $S$, the source generates an incident acoustic pressure field $p_{\textrm{inc}}(\mathbf{r},t)$ corresponding to free propagation in the homogeneous medium. In the presence of the object, this field induces a scattered field $p_{\textrm{scat}}(\mathbf{r},t)$ due to reflection from $S$. 
Our objective is to derive $p_{\textrm{scat}}(\mathbf{r},t)$ and thereby obtain the total field $p_{\textrm{tot}}(\mathbf{r},t) = p_{\textrm{inc}}(\mathbf{r},t) + p_{\textrm{scat}}(\mathbf{r},t)$ for all $\mathbf{r}$ and $t$.
Here, $t$ denotes the observation (arrival) time. The relation between $t$ and $t_e$ is discussed in Sec. \ref{sec:TDPO_theory}.
Before deriving the time-domain formulation, we briefly revisit the well-established frequency-domain PO solution.

\subsection{Frequency-domain PO analytical formulation}  \label{sec:freq}

Here it is assumed that all components in Fig.~\ref{fig:scat_field_scheme} are stationary, that the source generates a purely harmonic signal, and that the steady-state harmonic regime has been reached. Under these conditions, a frequency-domain formulation can be used with an assumed time dependence $e^{\mathrm{j}\omega t}$, which is suppressed for brevity.
The pressure field scattered by the object and observed at the field point $\mathbf{r}$ then becomes \cite{ColtonKress2019}
\begin{equation} \label{eq:p_scat}
    \widetilde{p}_{\textrm{scat}}(\mathbf{r};\omega)
    =
    \int_{S}
    \widetilde{p}_{\textrm{tot}}(\mathbf{r}';\omega)
    \frac{\partial G(\mathbf{R};\omega)}{\partial \widehat{\mathbf{n}}'}
    \, \mathrm{d}s'.
\end{equation}
Here, $\widetilde{p}_{\mathrm{tot}}(\mathbf{r}';\omega)$ denotes the total field on the object surface, and $\mathrm{d}s'$ is the differential surface element at $\mathbf{r}'$. The 3D free-space Green's function $G$, evaluated at a general observation position $\mathbf{r}$ for a source located at the origin, is defined as
\begin{equation}
\label{eq:Green_q}
G(\mathbf{r};\omega)
=
\frac{e^{-\mathrm{j}kr}}
{4\pi r},
\end{equation}
where $k=\omega/c$ is the wavenumber and $r=|\mathbf{r}|$.
Under the PO approximation, whose validity is examined in Sec.~\ref{sec:sim_fd_sphere}, the total field on the illuminated surface is approximated as
\begin{equation} \label{eq:PO_def}
    \widetilde{p}_{\textrm{tot}}(\mathbf{r}';\omega)\approx2\widetilde{p}_{\textrm{inc}}(\mathbf{r}';\omega),
\end{equation}
while on the shadowed part of $S$ it is set to zero, which restricts the integration in \eqref{eq:p_scat} to $S_{\textrm{lit}}$. 
The incident field, i.e., the free-space response of a source at $\mathbf{r}_e$, at a general observation point $\mathbf{r}$, is given by
\begin{equation} \label{eq:p_inc}
    \widetilde{p}_{\textrm{inc}}(\mathbf{r};\omega)=G(\mathbf{R}_e;\omega)\widetilde{q}_e,
\end{equation}
where 
$\widetilde{q}_e$ is the complex amplitude of the harmonic source signal. On the surface, it is evaluated at $\mathbf{R}'_e $  
as $\widetilde{p}_{\textrm{inc}}(\mathbf{r}';\omega)=G(\mathbf{R}'_e;\omega)\widetilde{q}_e$.
Substituting \eqref{eq:PO_def}--\eqref{eq:p_inc} into \eqref{eq:p_scat} and differentiating the Green's function \eqref{eq:Green_q} with respect to $\widehat{\mathbf{n}}'$, and denoting $\widehat{\mathbf{R}}=\mathbf{R}/R$, the scattered pressure field takes the form
\begin{equation}  \label{eq:p_scat_final}
    \widetilde{p}_{\textrm{scat}}(\mathbf{r};\omega)=\int_{S_{\textrm{lit}}}\frac{2\widetilde{q}_e e^{-\mathrm{j}kR_e'}}{4\pi R_e'}\left(\mathrm{j}k+\frac{1}{R}\right)\frac{e^{-\mathrm{j}kR}\,\widehat{\mathbf{R}}\cdot \widehat{\mathbf{n}}'}{4\pi R}\mathrm{d}s'.
\end{equation}

\subsection{Time-dependent PO analytical formulation} \label{sec:TDPO_theory}

The time-domain case differs from the frequency-domain scenario in two respects: (i) the monopole source time dependence $q_e(t_e)$ may be an arbitrary function, rather than a purely harmonic one, and (ii) the source may move continuously along a trajectory $\mathbf{r}_e(t_e)$. The implications of these two differences for the derivation are addressed next.

We first consider the free-space field \eqref{eq:p_inc}. For a constant distance $r$ from the origin, the transition to the time domain results in a fixed temporal shift, expressed via the inverse Fourier transform of the frequency-domain Green's function \eqref{eq:Green_q}, as $g(r,t)=\delta(t-r/c)/4\pi r$. Convolving this time domain Green's function with the signal of a stationary source at $\mathbf{r}_e$ yields $p_{\textrm{inc}}(\mathbf{r},t)=q_e(t_e)/4\pi R_e$, with $t_e=t-R_e/c$. 
For a moving source, $t_e$ and $t$ are related by
\begin{equation}  \label{eq:t_e}
    t_e=t-\frac{R_e(t_e)}{c}.
\end{equation}
For a point monopole of strength $q_e$ carried along $\mathbf{r}_e$ at a subsonic speed, $|\mathbf{v}_e|<c$, the free-space field is \cite{DowlingFfowcsWilliams1983,MorseIngard1968,casalino2003advanced}
\begin{equation} \label{eq:p_inc_time}
    p_{\textrm{inc}}(\mathbf{r},t)=\frac{q_e(t_e)}{4\pi R_e(t_e)\left[1+\mu_r(t_e)\right]},
\end{equation}
where
\begin{equation} \label{eq:mu_r}
    \mu_r(t_e)=\frac{1}{c}\frac{\partial \mathbf{R}_e(t_e)}{\partial t_e}\cdot \widehat{\mathbf{R}}_e(t_e)=-\frac{\mathbf{v}_e(t_e)\cdot \widehat{\mathbf{R}}_e(t_e)}{c}
\end{equation}
is the directional Mach number, namely the rate of change of the source-to-observer distance $R_e$ normalized by $c$, with $\widehat{\mathbf{R}}_e=\mathbf{R}_e/R_e$ and $\mathbf{v}_e=\mathrm{d}\mathbf{r}_e/\mathrm{d}t_e$ the source velocity. The factor $1/(1+\mu_r)=\mathrm{d}t_e/\mathrm{d}t$, which arises from the Jacobian of the retarded-time relation \eqref{eq:t_e}, is the Doppler amplitude factor, which accompanies the Doppler frequency shift contained in $q_e(t_e)$. Since $-\mu\leq\mu_r\leq\mu$ for a source with Mach number $\mu=|\mathbf{v}_e|/c$, the field is amplified ahead of the source ($\mu_r<0$) and attenuated behind it ($\mu_r>0$), by up to $1/(1\mp\mu)$. For a stationary source, the factor equals unity.

A similar treatment applies to the exponential terms in the scattered-field expression \eqref{eq:p_scat_final}. It is then convenient to express $t_e$ in terms of $t'$, the time at which the wave emitted from $\mathbf{r}_e(t_e)$ reaches the point $\mathbf{r}' \in S$. This incidence time is related to the observation time at $\mathbf{r}$ in the ROI via
\begin{equation} \label{eq:t_e_t'}
    t_e=t'-\frac{R_e'(t_e)}{c}, \qquad t'=t-\frac{R}{c}.
\end{equation}
Combining this with the fact that multiplication by $\mathrm{j}k$ in the frequency domain corresponds to $c^{-1}$ times a time derivative in the time domain, and noting that within the scattering integral this derivative must be taken with respect to the incidence time $t'$, the scattered field takes the form
\begin{equation}  \label{eq:p_scat_analytic}
    p_{\textrm{scat}}(\mathbf{r},t)=\frac{2}{4\pi}\int_{S_{\textrm{lit}}}\left[\frac{1}{c}\frac{\partial p_{\textrm{inc}}}{\partial t'}+\frac{p_{\textrm{inc}}}{R}\right]_{t'=t-R/c}\frac{\widehat{\mathbf{R}}\cdot\widehat{\mathbf{n}}'}{R}\,\mathrm{d}s',
\end{equation}
where the factor 2 is that of the PO surface field \eqref{eq:PO_def}, and $p_{\textrm{inc}}(\mathbf{r}',t')=q_e(t_e)/\{4\pi R'_e(t_e)[1+\mu_r'(t_e)]\}$ is the incident field \eqref{eq:p_inc_time} at the surface point, with $\mu_r'(t_e)$ given by \eqref{eq:mu_r}, both with $\mathbf{R}_e$ replaced by $\mathbf{R}'_e$, and $t_e$ related to $t'$ by \eqref{eq:t_e_t'}.

For a stationary source, $\mu_r'=0$ and $R'_e(t_e)=R'_e$, and the derivative in \eqref{eq:p_scat_analytic} simply reads $\partial p_{\textrm{inc}}/\partial t'=(\partial q_e/\partial t_e)/4\pi R'_e$.
However, for a moving source, the chain rule gives
\begin{multline}  \label{eq:p_inc_dot}
    \frac{\partial p_{\textrm{inc}}}{\partial t'}=\frac{1}{4\pi R_e'(t_e)\left[1+\mu_r'(t_e)\right]}\frac{\partial t_e}{\partial t'}\\
    \times\left[\frac{\partial q_e(t_e)}{\partial t_e}-\frac{c\, q_e(t_e)}{R_e'(t_e)}\mu_r'(t_e)-\frac{q_e(t_e)}{1+\mu_r'(t_e)}\frac{\partial \mu_r'(t_e)}{\partial t_e}\right],
\end{multline}
where
\begin{equation}  \label{eq:chain}
    \frac{\partial t_e}{\partial t'}=\frac{1}{1+\mu_r'(t_e)}
\end{equation}
is obtained by the same chain rule using the definitions in \eqref{eq:t_e_t'}. Differentiating \eqref{eq:mu_r} gives $\partial\mu_r'/\partial t_e=c(\mu^2-\mu_r'^2)/R'_e-\mathbf{a}_e\cdot\widehat{\mathbf{R}}'_e/c$, with $\mathbf{a}_e=\mathrm{d}\mathbf{v}_e/\mathrm{d}t_e$ the source acceleration and $\widehat{\mathbf{R}}'_e=\mathbf{R}'_e/R'_e$.
Equation \eqref{eq:p_scat_analytic}, with the derivative \eqref{eq:p_inc_dot}--\eqref{eq:chain}, forms the complete TDPO solution.

\subsection{Fock correction for a moving source} \label{sec:fock_theory}

The PO surface field \eqref{eq:PO_def} drops abruptly from $2\widetilde{p}_{\textrm{inc}}$ to zero at the shadow boundary, formed by the grazing points $\mathbf{r}'_g$ at which the rays from the source are tangent to $S$. For a smooth convex rigid object, Fock's theory \cite{Fock1965,Pathak1980} replaces this step by a continuous transition over the penumbra (the Fock region), a band of the surface around the boundary that narrows as the frequency increases. On $S_{\textrm{lit}}$, the factor 2 of \eqref{eq:PO_def} becomes Fock's factor $F_{\textrm{lit}}(\xi_{\textrm{lit}})$, and the shadowed part $S_{\textrm{sh}}$ is reached by creeping waves, which leave $\mathbf{r}'_g$ along the surface, travel the arc length $s$, and are weighted by $\mathcal{D}F_{\textrm{sh}}(\xi_{\textrm{sh}})$. Here, $\xi$ measures the distance from the boundary in penumbra widths, and $\mathcal{D}$ accounts for the spreading of the creeping waves. The frequency-domain form, which is well established, is summarized in Appendix~\ref{app:fock_fd} and Fig.~\ref{fig:fock_geometry} therein.

Since $\xi\propto\omega^{1/3}$, the Fock factors depend on frequency, and in the time domain the multiplication by them becomes a convolution with their inverse Fourier transforms, which differ from one surface point to another. For a sinusoidal signal $q_e(t_e)=\sin(\omega_0t_e)$, the convolution reduces to a multiplication by the factor at the frequency received at the surface point, which scales the sine by its magnitude and shifts its phase by its argument. The surface points then radiate the modified signals
\begin{equation} \label{eq:q_fock}
\begin{aligned}
q^{\textrm{lit}}_e(t_e)&=\tfrac{1}{2}\,|F_{\textrm{lit}}|\sin\!\left(\omega_0t_e+\arg F_{\textrm{lit}}\right),\\
q^{\textrm{sh}}_e(t_e)&=\tfrac{1}{2}\,\mathcal{D}\,|F_{\textrm{sh}}|\sin\!\left(\omega_0t_e+\arg F_{\textrm{sh}}\right),
\end{aligned}
\end{equation}
on $S_{\textrm{lit}}$ and $S_{\textrm{sh}}$, respectively, with $F_{\textrm{lit}}$ evaluated at the received frequency $\omega_0/(1+\mu_r')$. The correction changes only the surface field in \eqref{eq:p_scat_analytic}. On $S_{\textrm{lit}}$, $q_e$ is replaced by $q^{\textrm{lit}}_e$ in \eqref{eq:p_inc_dot}. The integration also extends to $S_{\textrm{sh}}$, where the creeping path $R'_g+s$, with $R'_g=|\mathbf{r}'_g-\mathbf{r}_e|$, replaces $R'_e$ in the delay, and therefore in the Doppler amplitude factor, whereas the amplitude retains $1/R'_g$. There, $p_{\textrm{inc}}$ and its derivative are replaced by
\begin{align}
    p^{\textrm{sh}}(\mathbf{r}',t')&=\frac{q^{\textrm{sh}}_e(t_e)}{4\pi R'_g(t_e)\left[1+\mu'_s(t_e)\right]}, \label{eq:p_sh}\\
    \frac{\partial p^{\textrm{sh}}}{\partial t'}&=\frac{\dot{q}^{\textrm{sh}}_e-q^{\textrm{sh}}_e\dot{R}'_g/R'_g-q^{\textrm{sh}}_e\dot{\mu}'_s/\left(1+\mu'_s\right)}{4\pi R'_g\left(1+\mu'_s\right)^2}, \label{eq:p_sh_dot}
\end{align}
where $\mu'_s=c^{-1}\mathrm{d}(R'_g+s)/\mathrm{d}t_e$ is the directional Mach number of the creeping path, the dots on $q^{\textrm{sh}}_e$, $R'_g$ and $\mu'_s$ denote $\mathrm{d}/\mathrm{d}t_e$, $\mathcal{D}F_{\textrm{sh}}$ is taken at the received frequency $\omega_0/(1+\mu'_s)$, and $t'=t_e+(R'_g+s)/c$ replaces \eqref{eq:t_e_t'}. Unlike in \eqref{eq:p_inc_dot}, the second term of the numerator contains the rate of $R'_g$, which differs from that of the delay path. The geometry ($\theta_i$, $\mathbf{r}'_g$, $s$, $\mathcal{D}$) is evaluated at the emission time, and the factors, which vary slowly for subsonic motion, are held constant in the derivative. On the shadow boundary, $F_{\textrm{lit}}(0)=F_{\textrm{sh}}(0)$, $\mathcal{D}=1$ and $R'_g=R'_e$, so that the surface field varies continuously across it. With $F_{\textrm{lit}}=2$ and no shadowed contribution, \eqref{eq:p_scat_analytic} is recovered, and, for a stationary harmonic source, the frequency-domain Fock result of Appendix~\ref{app:fock_fd}.

\section{Numerical Evaluation}  \label{sec:numerical}

In \eqref{eq:p_scat_analytic}--\eqref{eq:chain}, the emission time is related to the incidence time only implicitly, through \eqref{eq:t_e_t'}. This relation is handled by two approaches, termed \emph{backward} and \emph{forward}, which are derived first for the free-space field and then extended to the scattered field.

\subsection{Numerical evaluation of the free-space field}  \label{sec:num_free}

In the backward approach, the field is expressed in terms of the observation time $t$. 
With $t_e(t)$ obtained from \eqref{eq:t_e}, the free-field response \eqref{eq:p_inc_time} reads
\begin{equation} \label{eq:p_inc_time_back}
    p_\textrm{inc}(\mathbf{r},t)=\frac{q_e(t_e(t))}{4\pi R_e(t_e(t))\left[1+\mu_r(t_e(t))\right]}.
\end{equation}
This formulation yields the field directly at any prescribed observation time $t$ and location $\mathbf{r}$, once $t_e(t)$ is known, either in closed form or numerically.
As an illustrative example, consider a source moving at constant speed $v_e$ along a straight line segment that starts at $\mathbf{r}_s$ and is oriented along $\widehat{\mathbf{n}}_e$, such that $\mathbf{r}_e(t_e)=\mathbf{r}_s + \mathbf{v}_e t_e $, where $\mathbf{v}_e=v_e\widehat{\mathbf{n}}_e$, $v_e=\mu c$, and $\mu\in(0,1)$ is the Mach number. Substitution into \eqref{eq:t_e} yields
\begin{equation} \label{eq:f(t)}
    t_e(t)= \frac{t - \gamma
    - \sqrt{
    \mu^2 t^2
    - 2 \gamma  t
    + \gamma ^2
    + \frac{1 - \mu^2}{c^2} R_s^2
    }}{1 - \mu^2}
\end{equation}
where $R_s=|\mathbf{R}_s |$,  $\mathbf{R}_s=\mathbf{r}-\mathbf{r}_s$, and $\gamma = \frac{\mu}{c} \mathbf{R}_s \cdot \widehat{\mathbf{n}}_e$. Even for this simple trajectory, the expression for $t_e(t)$ is rather lengthy.

For a general trajectory, including one defined by a collection of sampled points, $t_e(t)$ is obtained numerically by solving $t_e+R_e(t_e)/c-t=0$ for $t_e$. For a subsonic source, $|\mu_r|<1$, so the derivative of the left-hand side with respect to $t_e$, $1+\mu_r(t_e)$, is positive. The left-hand side therefore increases strictly with $t_e$, and the solution is unique and lies in the interval $t-R_{\max}/c\leq t_e\leq t$, with $R_{\max}$ the largest source-to-observer distance. It can be found by bisection within this interval or by Newton iterations, in which each estimate of $t_e$ is corrected by $-[t_e+R_e(t_e)/c-t]/[1+\mu_r(t_e)]$, whose denominator is the Doppler factor of \eqref{eq:p_inc_time_back}.
Numerically, the observation time $t$ is advanced sequentially, and $p_\textrm{inc}(\mathbf{r},t)$ is evaluated accordingly.

In the forward approach, the solution is expressed in terms of the emission time $t_e$, such that the field in \eqref{eq:p_inc_time} becomes
\begin{equation} \label{eq:p_inc_te}
p_\textrm{inc}\!\left(\mathbf{r},\, t_e + \frac{R_e(t_e)}{c}\right)
= \frac{q_e(t_e)}{4\pi R_e(t_e)\left[1+\mu_r(t_e)\right]}.
\end{equation}
Formulation \eqref{eq:p_inc_te} is explicit in $t_e$: each emission instant is mapped directly to its arrival time $t_e+R_e(t_e)/c$, so that \eqref{eq:t_e} need not be solved, for any source trajectory $\mathbf{r}_e(t_e)$. However, the corresponding observation times are generally nonuniform with respect to a prescribed time grid.

Numerically, the emission time is sampled as $t_e^{(i)}=i\Delta t$, and each sample is assigned to its arrival time $t^{(i)}=t_e^{(i)}+R_e^{(i)}/c$, with the superscript $(i)$ denoting evaluation at $t_e^{(i)}$. For a subsonic source, $t^{(i)}$ increases monotonically with $i$, with spacing $(1+\mu_r)\Delta t$, compressed ahead of the source and dilated behind it. The field on the uniform time grid is obtained by interpolating these samples at each observation point. Since the Doppler shift scales the received period by the same factor $1+\mu_r$, each received period contains $T/\Delta t$ samples, with $T$ the period of $q_e$, as for a stationary source, so that the interpolation error does not depend on the motion. Since $\Delta t$ is also the step of the uniform time grid, it must resolve the shortest received period, $(1-\mu)T$ ahead of the source, and not only $T$.

\begin{figure}[!tb]
    \centering
    \includegraphics[width = 0.32\textwidth]{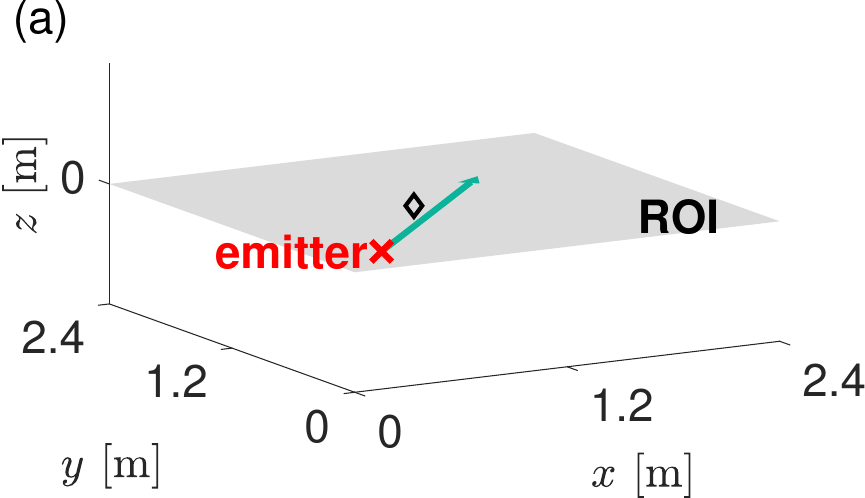} \\
    \includegraphics[scale = 0.42]{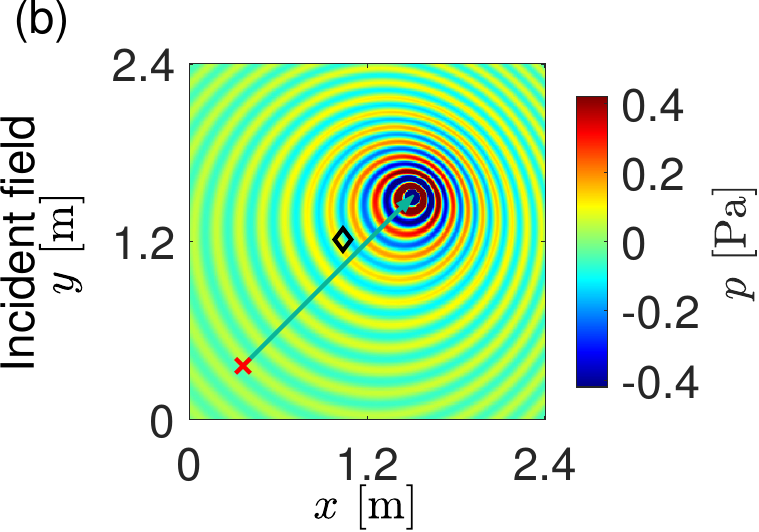} \\
    \includegraphics[scale = 0.40]{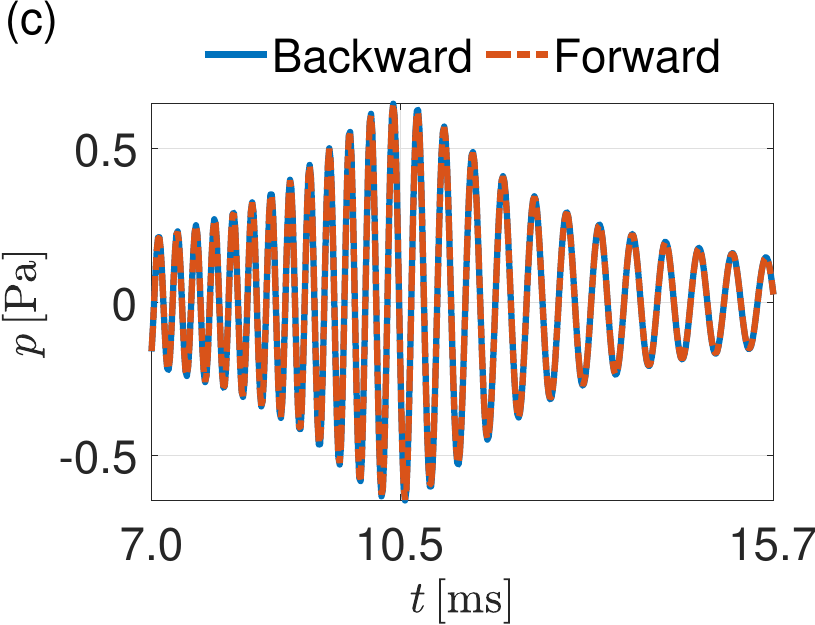}
\caption{\textbf{Free-space field simulation}. (a) Setup: the ROI (gray), the source's initial position (red cross) and its trajectory (turquoise arrow), and the observation point (black diamond). (b) Snapshot at $t=15.7$~ms; the color scale is clipped at $\pm0.42$~Pa, which the field exceeds near the source, where it grows as $1/R_e$. (c) Response at the observation point, $(x,y)=(1.03,\,1.21)$~m, passed by the source at $t=10.5$~ms, up to the snapshot instant; the relative rms difference between the two approaches over the ROI and all time steps, excluding the clipped region, is $e_{\mathrm{rms}}=0.90\%$.}
    \label{fig:forward_sim}
\end{figure}

The setup is shown in Fig.~\ref{fig:forward_sim}(a). A sinusoidal source of frequency $f_0=2730$~Hz in air, with $c=343$~m/s, so that $\lambda=0.126$~m and $T=1/f_0\approx 0.37$~ms, starts at $(x,y)=(0.36,\,0.36)$~m and moves at Mach number $\mu=0.3$ along a straight line at $45^\circ$ to the $x$-axis. The field is computed on the plane $z=0$, which contains the trajectory, with the spatial step $\Delta x=\lambda/20$ and the time step $\Delta t=\Delta x/c=T/20\approx 18.3~\mu$s, i.e., 14 samples per shortest period $(1-\mu)T$. Figure~\ref{fig:forward_sim}(b) shows a snapshot of the field computed by the forward approach, and Fig.~\ref{fig:forward_sim}(c) compares the forward and backward approaches at a single observation point.
In the next subsection, we describe the numerical evaluation of the scattered field using both approaches.

\begin{figure*}[!tb]
    \centering   
          \includegraphics{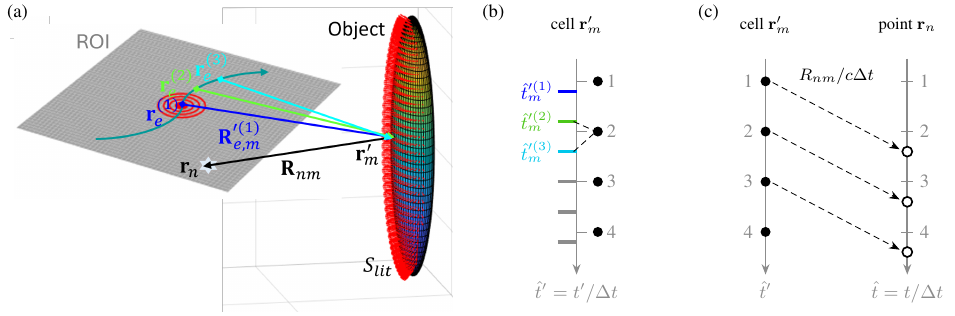}
  \caption{\textbf{Two-stage evaluation of the scattered field.}
(a) Propagation from the source positions $\mathbf{r}_e^{(i)}$ of three successive emission samples to the surface cell $\mathbf{r}'_m$, and from $\mathbf{r}'_m$ to the observation point $\mathbf{r}_n$.
(b) Stage~1: normalized incidence times $\hat{t}'^{(i)}_m=t'^{(i)}_m/\Delta t$ at $\mathbf{r}'_m$, spaced by $1+\mu'^{(i)}_{r,m}$, and the uniform samples (dots) interpolated from them; each uniform sample lies between one pair of successive incidence times (dashed lines).
(c) Stage~2: the uniform samples of cell $m$ reach $\mathbf{r}_n$ delayed by the fixed $R_{nm}/c$ (open circles).}
    \label{fig:scat_field_forward}
\end{figure*}

\subsection{Numerical evaluation of the scattered field}  \label{sec:num_scat}

The scattered field \eqref{eq:p_scat_analytic} combines two propagation paths: from the source to the surface point $\mathbf{r}'$, and from $\mathbf{r}'$ to the observation point $\mathbf{r}$, as illustrated in Fig.~\ref{fig:scat_field_forward}(a). By \eqref{eq:t_e_t'}, only the first path depends on the source motion. Since the object and the ROI are stationary, the second path contributes the fixed delay $R/c$. The integrand of \eqref{eq:p_scat_analytic} therefore separates into the incident field at the surface point, which depends on the incidence time $t'$ alone, and a time-independent factor and delay $R/c$ for each pair of a surface point and an observation point.
The surface $S$ is discretized into cells centered at $\mathbf{r}'_m$ with areas $\Delta s_m$, and the ROI into points $\mathbf{r}_n$. With $\mathbf{R}'_{e,m}(t_e)=\mathbf{r}'_m-\mathbf{r}_e(t_e)$ and $\mathbf{R}_{nm}=\mathbf{r}_n-\mathbf{r}'_m$, the scattered field \eqref{eq:p_scat_analytic} reads
\begin{equation} \label{eq:p_scat_sum}
    p_{\textrm{scat}}(\mathbf{r}_n,t)=\frac{2}{4\pi}\sum_{m} w_{nm}\left[\frac{\dot{p}_{\textrm{inc},m}}{c}+\frac{p_{\textrm{inc},m}}{R_{nm}}\right],
\end{equation}
where the bracket is evaluated at the incidence time $t'=t-R_{nm}/c$, $w_{nm}=\Delta s_m\,\widehat{\mathbf{R}}_{nm}\cdot\widehat{\mathbf{n}}'_m/R_{nm}$ and $p_{\textrm{inc},m}(t')=p_{\textrm{inc}}(\mathbf{r}'_m,t')$. The derivative $\dot{p}_{\textrm{inc},m}=\partial p_{\textrm{inc},m}/\partial t'$ is given by \eqref{eq:p_inc_dot}--\eqref{eq:chain}, with $\mu'_{r,m}$ the directional Mach number toward the cell. The sum runs over the cells illuminated at the emission time, $\mathbf{R}'_{e,m}(t_e)\cdot\widehat{\mathbf{n}}'_m<0$, and, with the Fock correction, also over the shadowed cells, through \eqref{eq:p_sh}--\eqref{eq:p_sh_dot}. In \eqref{eq:p_inc_dot}, $\partial q_e/\partial t_e$, $\mathbf{v}_e$ and $\mathbf{a}_e$ are finite differences of the sampled source signal and trajectory; differencing the incident field at the cell instead would add impulses where the cell switches on and off with the illumination.

The weight $w_{nm}$ and the delay $R_{nm}/c$ depend only on the positions of the cell and the observation point, whereas the time dependence, the source motion and the illumination are contained in $p_{\textrm{inc},m}$.
Accordingly, \eqref{eq:p_scat_sum} is evaluated in two stages. The first maps the emission samples onto a uniform incidence-time grid at each surface cell $\mathbf{r}'_m$, and the second sums the contributions of all cells at each ROI point $\mathbf{r}_n$ with their fixed delays. Mapping each emission sample directly to its arrival time $t_e+(R'_{e,m}+R_{nm})/c$ at the ROI would instead require the nonuniform arrival times to be treated separately for every cell-observation pair.

In stage 1, $p_{\textrm{inc},m}$ and $\dot{p}_{\textrm{inc},m}$ are obtained for each surface cell on the uniform grid $t'_k=k\Delta t$. This is the free-space problem of Sec.~\ref{sec:num_free}, with the observation point replaced by $\mathbf{r}'_m$ and $t$ by $t'$, so that both approaches of that section apply.
In the forward approach, each emission sample $t_e^{(i)}=i\Delta t$ is mapped to the incidence time $t'^{(i)}_m=t_e^{(i)}+R'^{(i)}_{e,m}/c$, generally not a multiple of $\Delta t$, at which $p_{\textrm{inc},m}$ and $\dot{p}_{\textrm{inc},m}$ are evaluated from \eqref{eq:p_inc_time} and \eqref{eq:p_inc_dot}, or set to zero if the cell is not illuminated at $t_e^{(i)}$. Successive incidence times are spaced by $(1+\mu'^{(i)}_{r,m})\Delta t$, where $\mu'^{(i)}_{r,m}=(R'^{(i)}_{e,m}-R'^{(i-1)}_{e,m})/c\Delta t$ is the discrete counterpart of $\mu'_{r,m}$. When the source approaches the cell, $\mu'^{(i)}_{r,m}<0$ and several incidence times may fall within one sampling interval, as shown in Fig.~\ref{fig:scat_field_forward}(b).

For a subsonic source, the incidence times increase monotonically, so that each grid sample $t'_k$ lies between exactly one pair of successive incidence times and is interpolated from it. The other incidence times within an interval are redundant, provided that $\Delta t$ resolves $(1-\mu)T$, as in Sec.~\ref{sec:num_free}. 

Stage 1 is independent of the ROI: for $N_e$ emission samples and $N_c$ cells, it consists of $N_e\times N_c$ array operations and an interpolation per cell. These are applied to the whole record when the trajectory is known in advance, or in a loop over $t_e^{(i)}$, vectorized over the cells, when it becomes available only during the motion, at the same cost.

In stage 2, \eqref{eq:p_scat_sum} is a delay-and-sum of the uniformly sampled incident fields at the cells. For a stationary ROI, each pair has a fixed delay $R_{nm}/c$ [Fig.~\ref{fig:scat_field_forward}(c)], generally not a multiple of $\Delta t$. It is applied in the frequency domain, as the phase factor $e^{-\mathrm{j}\omega R_{nm}/c}$ on the discrete Fourier transforms of $p_{\textrm{inc},m}$ and $\dot{p}_{\textrm{inc},m}$, zero-padded beyond the longest delay, so that the inverse transform gives $p_{\textrm{scat}}(\mathbf{r}_n,t_k)$ without a second interpolation in time. For each frequency in the signal band, the sum over the cells is one matrix--vector product for all ROI points. This form requires the whole record and is thus offline. 

Online, the delay is applied by fractional-delay interpolation in time, and the two stages are interleaved: at each step, stage 1 extends the incident fields at the cells by one sample, and stage 2 evaluates them at $t_k-R_{nm}/c$.
For a moving ROI point $\mathbf{r}_n(t)$, $w_{nm}$ and $R_{nm}$ depend explicitly on the observation time; stage 1 is unchanged, and stage 2 runs as a loop over $t_k$ with the same fractional-delay interpolation.

In the backward approach, stage 1 is carried out by finding, for each cell and each grid sample $t'_k$, the emission time $t_e$ as the solution of $t_e+R'_{e,m}(t_e)/c-t'_k=0$, as in Sec.~\ref{sec:num_free}, and the illumination of the cell is checked at that emission time.
For the straight-line trajectory, $\mathbf{R}'_{e,m}(t_e)=\mathbf{R}'_{s,m}-v_e t_e\widehat{\mathbf{n}}_e$ with $\mathbf{R}'_{s,m}=\mathbf{r}'_m-\mathbf{r}_s$, and the solution is given by \eqref{eq:f(t)} with $t$, $R_s$ and $\gamma$ replaced by $t'_k$, $R'_{s,m}$ and $\gamma'_m=\mu\,\mathbf{R}'_{s,m}\cdot\widehat{\mathbf{n}}_e/c$. In this case, $\mu'_{r,m}(t_e)=\mu(v_e t_e-\mathbf{R}'_{s,m}\cdot\widehat{\mathbf{n}}_e)/R'_{e,m}(t_e)$ and $\partial\mu'_{r,m}/\partial t_e=c(\mu^2-\mu'^2_{r,m})/R'_{e,m}$.

For a general trajectory, the solution is found numerically, and since $R_{nm}$ does not enter the equation, stage 2 is the same as in the forward approach. The backward approach, however, requires an iterative solution for each cell and grid sample, with the trajectory and the source signal interpolated between their samples, whereas the forward approach uses only the samples. The latter is thus used in the simulations below.

\section{Simulations}   \label{sec:sim}

The method is first examined for a stationary source and a sphere in the frequency domain, where PO is established and the exact series solution is known. TDPO is then validated against the exact solution for a sphere and a moving source, and finally applied to an ellipsoid and a moving source, for which no exact solution is available. For comparison, FE simulations are also carried out in the frequency and time domains, as described in Appendix~\ref{app:fe}.

\begin{figure*}[!tb]
    \centering
    \includegraphics[width = 0.32\textwidth]{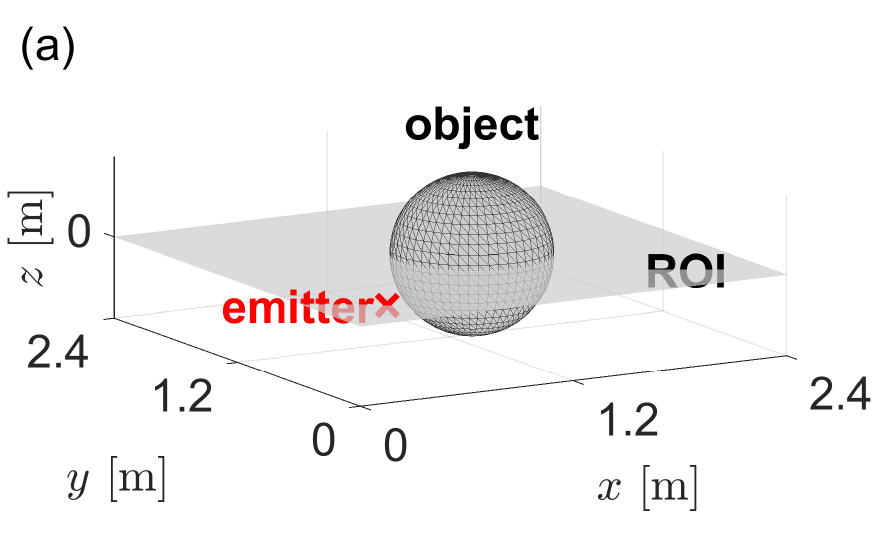}\hfill
    \includegraphics[width = 0.32\textwidth]{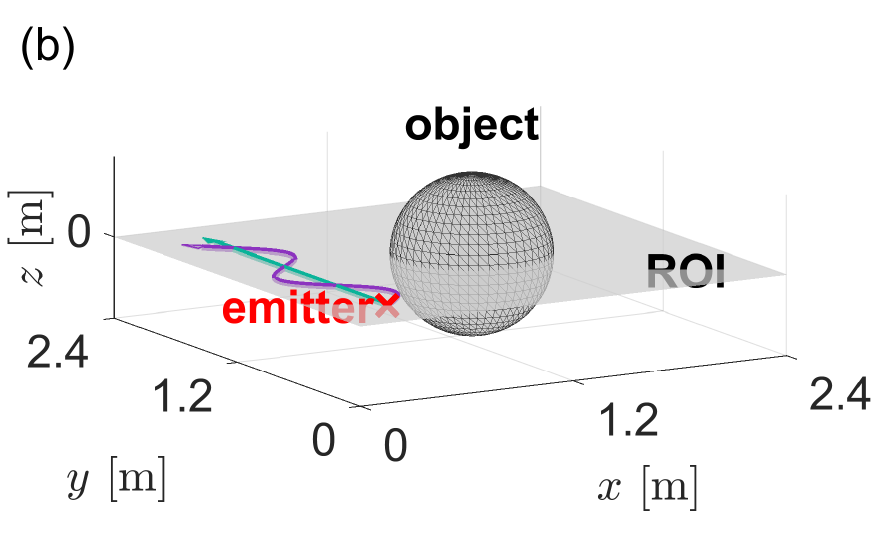}\hfill
    \includegraphics[width = 0.32\textwidth]{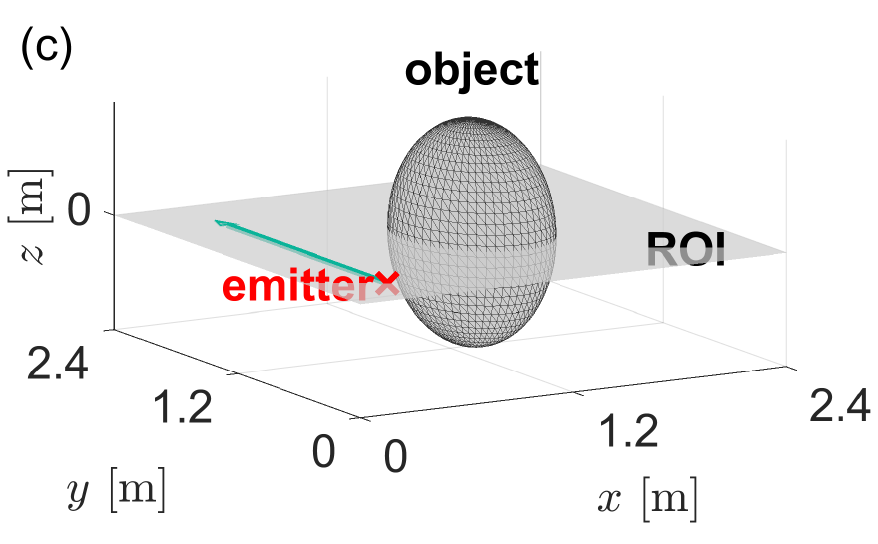}
  \caption{\textbf{Simulation setups.} (a) Stationary source and sphere (Sec.~\ref{sec:sim_fd_sphere}). (b) Moving source and sphere (Sec.~\ref{sec:sim_sphere}). (c) Moving source and ellipsoid (Sec.~\ref{sec:sim_ellipsoid}). Red cross: the source, or its initial position; turquoise arrow: the straight path; purple arrow: the curved path.}
    \label{fig:setups}
\end{figure*}

The simulation setups are shown in Fig.~\ref{fig:setups}. In all of them, the object is centered at $(x,y,z)=(1.32,\,1.2,\,0)$~m, the source is at, or starts from, $(0.36,\,0.36,\,0)$~m, and the scattered field is computed on the plane $z=0$, which contains the source and bisects the object. All simulations are conducted in air, the objects are acoustically rigid, and the monopole strength is $q_e=1\,\mathrm{kg/s^2}$. Unless stated otherwise, the PO surface cells, the FE mesh and the ROI grid are of size $\lambda/5$; in the time domain, the duration is $t_{\textrm{end}}=17.5$~ms, $\Delta t=T/20$, and the field is interpolated for display onto a grid of step $\lambda/20$.

\subsection{Stationary source and sphere} \label{sec:sim_fd_sphere}

\begin{figure*}[!tb]
    \centering
    \includegraphics[width = \textwidth]{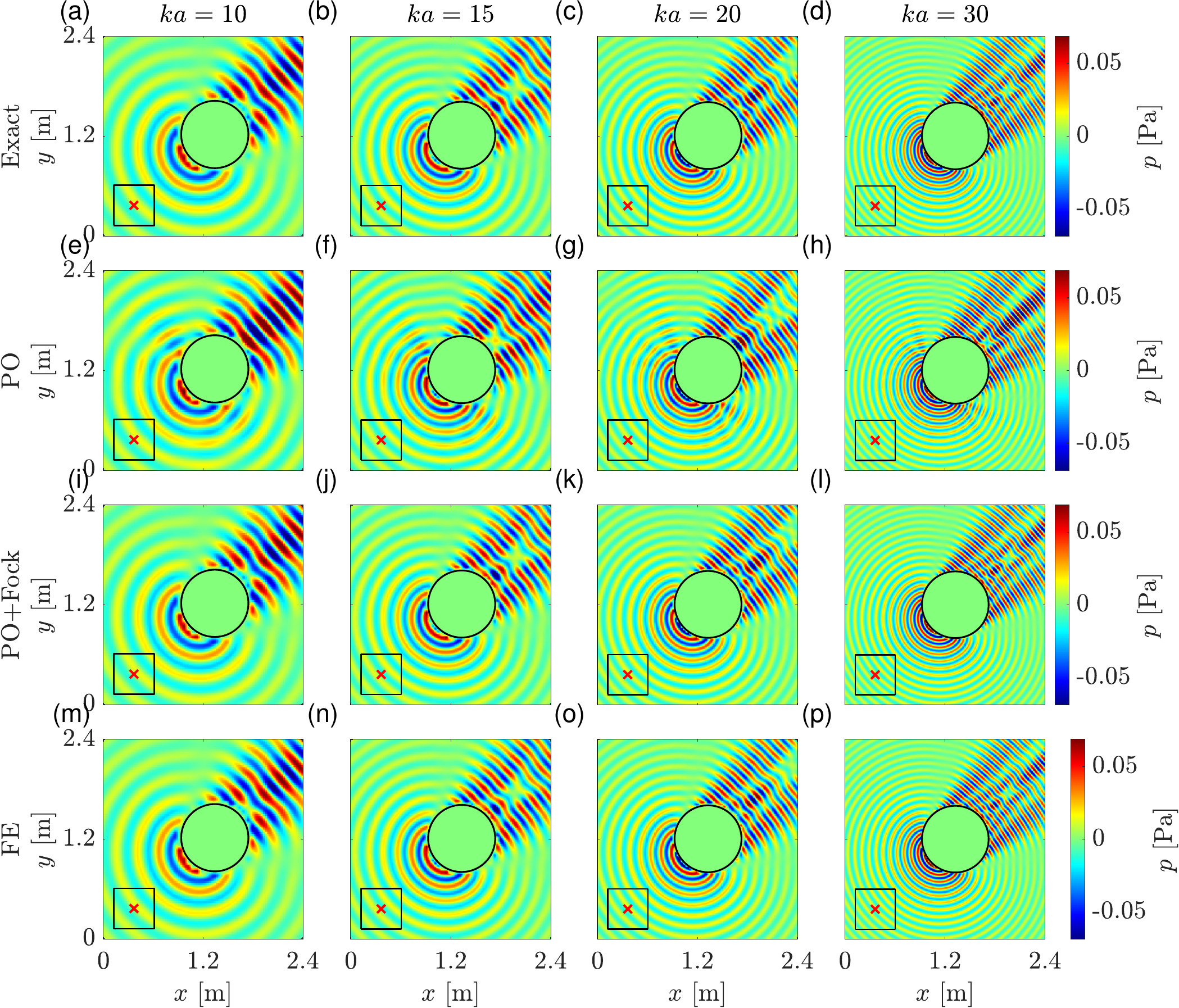}
  \caption{\textbf{Stationary source and sphere, frequency domain: scattered field.} Rows: exact solution, PO, PO with the Fock correction, and FE; columns: increasing $ka$. Instantaneous field at $t=0$ on the ROI. The red cross marks the source and the square the window around it.}
    \label{fig:fd_sphere}
\end{figure*}

\begin{figure*}[!tb]
    \centering
    \includegraphics[width = \textwidth]{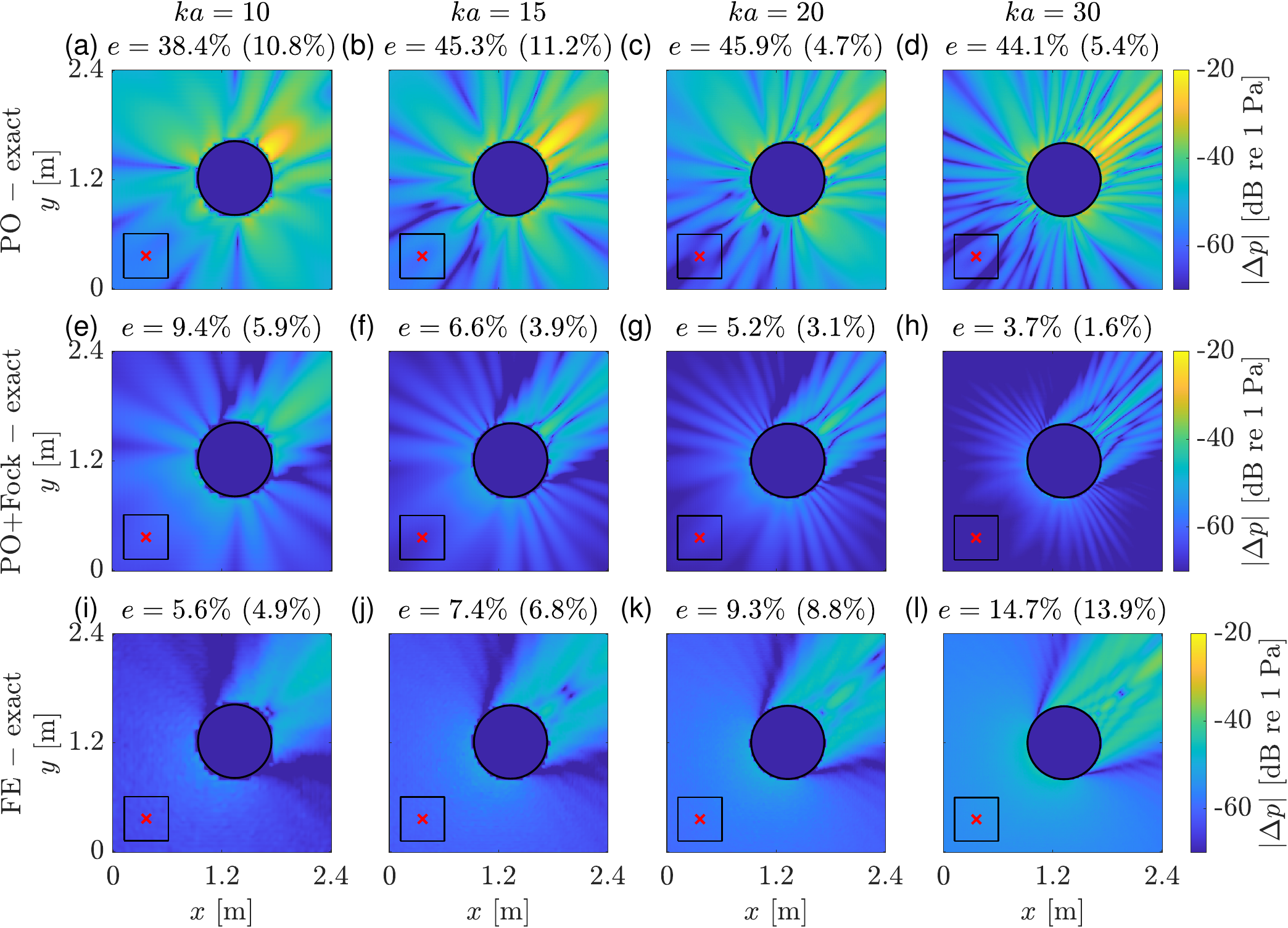}
  \caption{\textbf{Stationary source and sphere, frequency domain: errors.} Magnitude of the difference $|\Delta p|$ between the complex fields of Fig.~\ref{fig:fd_sphere}. Rows: PO, PO with the Fock correction, and FE, each against the exact solution; columns: increasing $ka$. Above each panel, $e=\|\Delta p\|/\|p_{\textrm{exact}}\|$ over the ROI and, in brackets, over the window.}
    \label{fig:fd_sphere_diff}
\end{figure*}

In the setup of Fig.~\ref{fig:setups}(a), a sphere of radius $a=0.4$~m is illuminated by a stationary source radiating a harmonic signal, for $ka=10$, 15, 20 and 30, i.e., $\lambda$ between 0.25~m and 0.084~m. The PO integral \eqref{eq:p_scat_final} is evaluated without and with the Fock correction of Appendix~\ref{app:fock_fd}, and the exact field is obtained from the series solution \eqref{eq:G_sphere} of Appendix~\ref{app:sphere_exact}. The error of a field $p$ is quantified by $e=\|p-p_{\textrm{exact}}\|/\|p_{\textrm{exact}}\|$ over the ROI points outside the sphere, and also over a $0.48\times0.48$~m$^2$ window centered on the source. Figure~\ref{fig:fd_sphere} shows the scattered fields, and Fig.~\ref{fig:fd_sphere_diff} the errors.

Over the whole ROI, which includes the shadow region behind the sphere, the PO error is 38.4\%--45.9\% and does not decrease with $ka$ [Fig.~\ref{fig:fd_sphere_diff}(a)--(d)]. PO truncates the surface field abruptly at the shadow boundary, which radiates spurious waves in all directions, seen as the radial streaks in Fig.~\ref{fig:fd_sphere_diff}(a)--(d), and it omits the creeping waves in the shadow; the largest PO error is in the forward direction. 

With the Fock correction, the surface field varies continuously across the penumbra and continues into the shadow, and the error decreases with $ka$, from 9.4\% at $ka=10$ to 3.7\% at $ka=30$ [Fig.~\ref{fig:fd_sphere_diff}(e)--(h)]. This decrease reflects the asymptotic nature of the correction for acoustically large objects: on a sphere, the penumbra $|\xi|\lesssim1$ has the half-width $a(2/ka)^{1/3}$ (Appendix~\ref{app:fock_fd}) \cite{Fock1965,Pathak1980}. The correction also reduces the error on the illuminated side of the sphere.

The FE error grows with $ka$, from 5.6\% at $ka=10$ to 14.7\% at $ka=30$, and is spread over the ROI, with the largest values in the forward direction [Fig.~\ref{fig:fd_sphere_diff}(i)--(l)]. Hence, for $ka\geq15$, PO with Fock is closer to the exact solution than FE. It is also faster by about three orders of magnitude, a fraction of a second to seconds per $ka$, against tens of seconds to minutes for each of the two FE solves.

In the window around the source, which corresponds to a monostatic configuration such as a sonar, the field is dominated by the reflection from the illuminated side, where PO is more accurate. There, the PO error roughly halves from $ka\leq15$ to $ka\geq20$, to about 5\% [Fig.~\ref{fig:fd_sphere_diff}(a)--(d)], and with the Fock correction it falls to 1.6\% at $ka=30$, below the FE error from $ka=15$ on [Fig.~\ref{fig:fd_sphere_diff}(e)--(l)]. Thus, while the Fock correction reduces the error over the ROI by a factor of 4 to 12, its effect near the source is moderate, and for receivers close to the source PO alone may suffice. This simplifies the computation, since only the lit cells, about a third of the surface cells here, contribute, and no creeping paths have to be traced over the shadowed part.

\subsection{Moving source and sphere} \label{sec:sim_sphere}

In the setup of Fig.~\ref{fig:setups}(b), the sphere of Sec.~\ref{sec:sim_fd_sphere} is illuminated by a source moving at Mach number $\mu=0.3$ along a straight line and radiating a sinusoidal signal with the frequency of Sec.~\ref{sec:num_free}, so that $ka=20$. The scattered field is computed by the two-stage scheme of Sec.~\ref{sec:num_scat}. The exact field \eqref{eq:p_scat_exact} superposes the fields scattered by a stationary monopole at the source position at each emission instant.

\begin{figure*}[!tb]
    \centering
    \includegraphics[width = \textwidth]{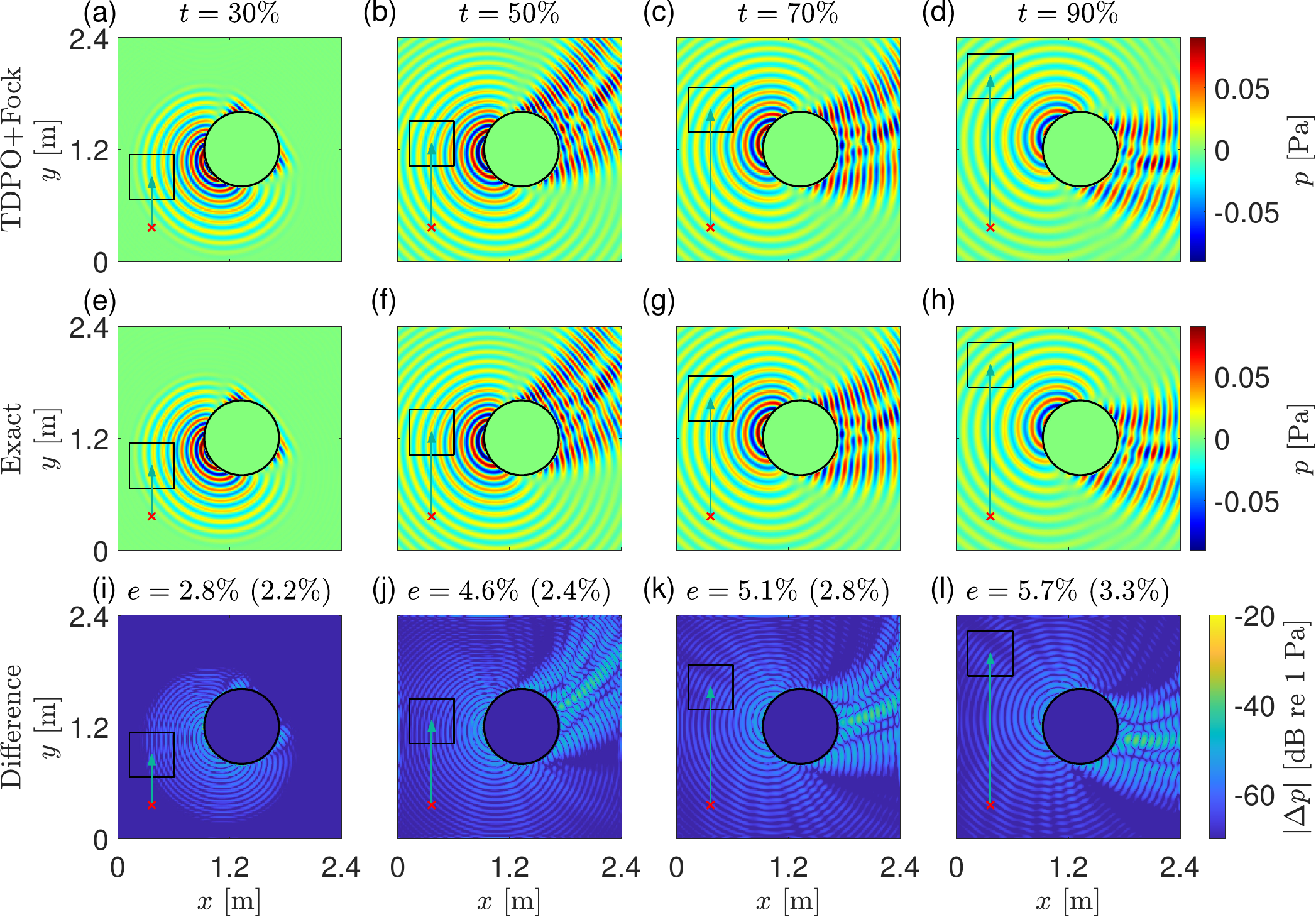}
  \caption{\textbf{Moving source and sphere, time domain.} Setup of Fig.~\ref{fig:setups}(b), $ka=20$. Rows: TDPO, the exact solution and their difference; columns: snapshots of the scattered field at increasing times, given as percentages of $t_{\textrm{end}}=17.5$~ms. The square marks the window around the source at the snapshot time. Above the bottom panels, $e$ is the relative rms error over the ROI and, in brackets, over the window.}
    \label{fig:sim_sphere}
\end{figure*}

Figure~\ref{fig:sim_sphere} compares TDPO with the Fock correction to the exact solution. As in Sec.~\ref{sec:sim_fd_sphere}, the largest differences are behind the sphere [Figs.~\ref{fig:sim_sphere}(k),(l) and \ref{fig:fd_sphere_diff}(e)--(h)]. When the scattered field fills the entire ROI, at $t=15.7$~ms, the error is 5.7\%, close to the 5.2\% obtained there at the same $ka$ [Figs.~\ref{fig:sim_sphere}(l) and \ref{fig:fd_sphere_diff}(g)]. Hence, at Mach number 0.3, TDPO retains the accuracy it has for a stationary source.

To demonstrate a broadband signal and a curved trajectory, the source radiates a train of equally spaced Ricker wavelets with central frequency $f_0=2730$~Hz, while moving at Mach number $\mu=0.3$ along the sinusoidal path of Fig.~\ref{fig:setups}(b) (purple). As this signal is broadband, the Fock correction, formulated in Sec.~\ref{sec:fock_theory} for a harmonic signal, is not applied here. Since the spectrum of a Ricker wavelet extends to about $2f_0$, the surface cells are of size $\lambda/10$, where $\lambda$ is the wavelength at $f_0$, so that at $2f_0$ they keep the resolution of $\lambda/5$ used for the sinusoidal source. The exact field is given by \eqref{eq:p_scat_exact} with the same source signal.

\begin{figure*}[!tb]
    \centering
    \includegraphics[width = \textwidth]{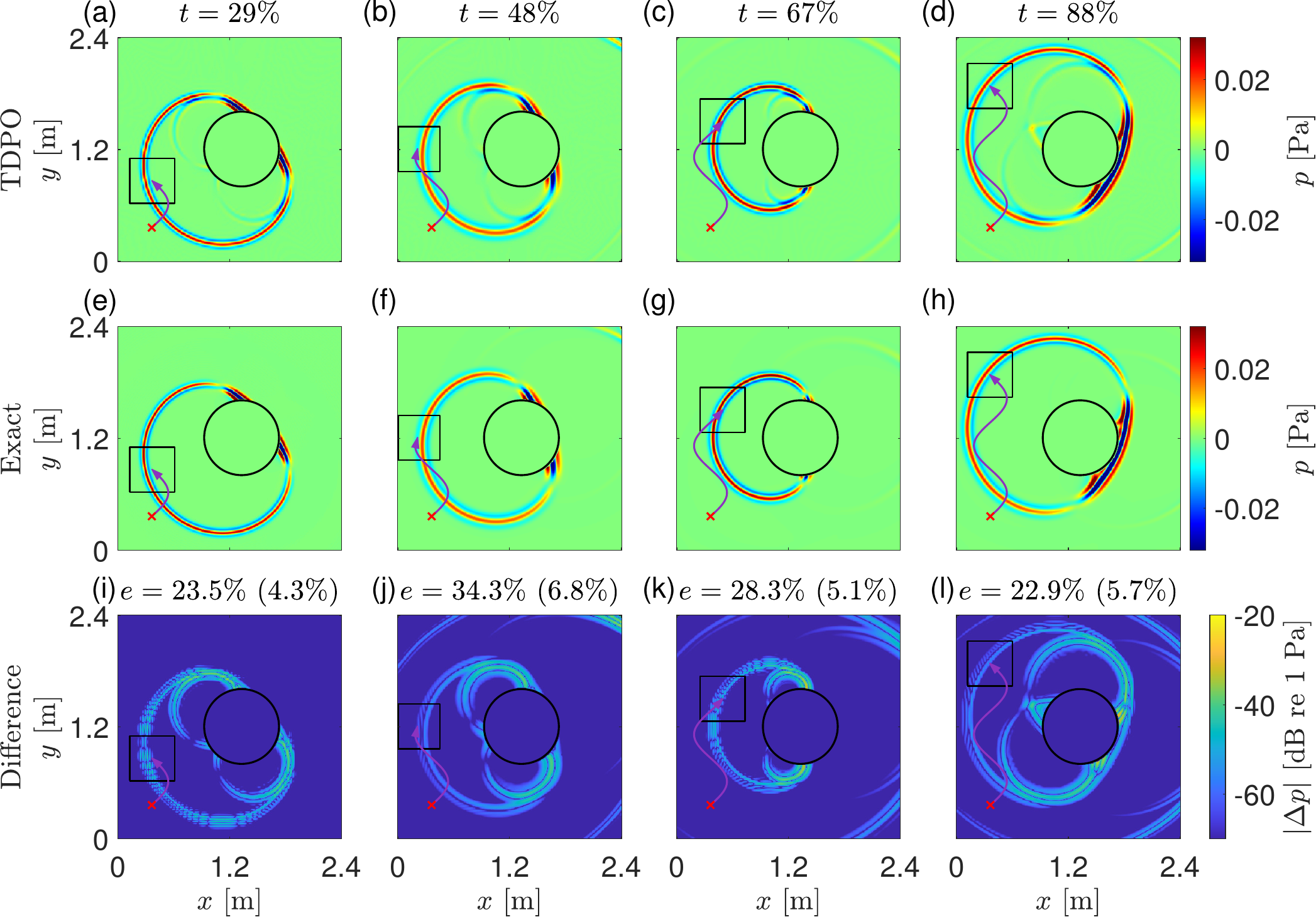}
  \caption{\textbf{Moving source on a curved path, broadband signal, and sphere.} Setup of Fig.~\ref{fig:setups}(b) with the curved path (purple). Rows, columns, markers and $e$ as in Fig.~\ref{fig:sim_sphere}, with TDPO without the Fock correction and times given as percentages of $t_{\textrm{end}}=21.0$~ms.}
    \label{fig:sim_ricker}
\end{figure*}

Figure~\ref{fig:sim_ricker} compares TDPO with the exact solution when the backscatter of successive wavelets of the train crosses the window around the source. Over the ROI, the error comes mostly from behind the sphere, where the creeping waves that the Fock correction would add are missing. In the window, where the backscatter dominates, the error is close to that of PO in Sec.~\ref{sec:sim_fd_sphere} at the same $ka$ [Figs.~\ref{fig:sim_ricker}(i)--(l) and \ref{fig:fd_sphere_diff}(c)]. Hence, for a broadband signal and a curved trajectory as well, TDPO retains near the source the accuracy it has for a stationary source.

\subsection{Moving source and ellipsoid} \label{sec:sim_ellipsoid}

In the setup of Fig.~\ref{fig:setups}(c), we consider an ellipsoid with semi-axes $a_e/\lambda=3$, $b_e/\lambda=4$ and $c_e/\lambda=4.5$, i.e., $ka_e\approx18.8$, $kb_e\approx25.1$ and $kc_e\approx28.3$, at the wavelength of Sec.~\ref{sec:num_free}. The source moves along the straight path at Mach number $\mu=0.2$, with the sinusoidal signal of Fig.~\ref{fig:sim_sphere}. On the ellipsoid, the radius of curvature varies along the surface, and the creeping paths and the spreading factor $\mathcal{D}$ of the Fock correction are obtained by tracing surface geodesics from the shadow boundary (Appendix~\ref{app:fock_fd}). Since no exact solution is available, TDPO with the Fock correction is compared with a transient FE simulation of the same setup, described in Appendix~\ref{app:fe}.

\begin{figure*}[!tb]
    \centering
    \includegraphics[width = \textwidth]{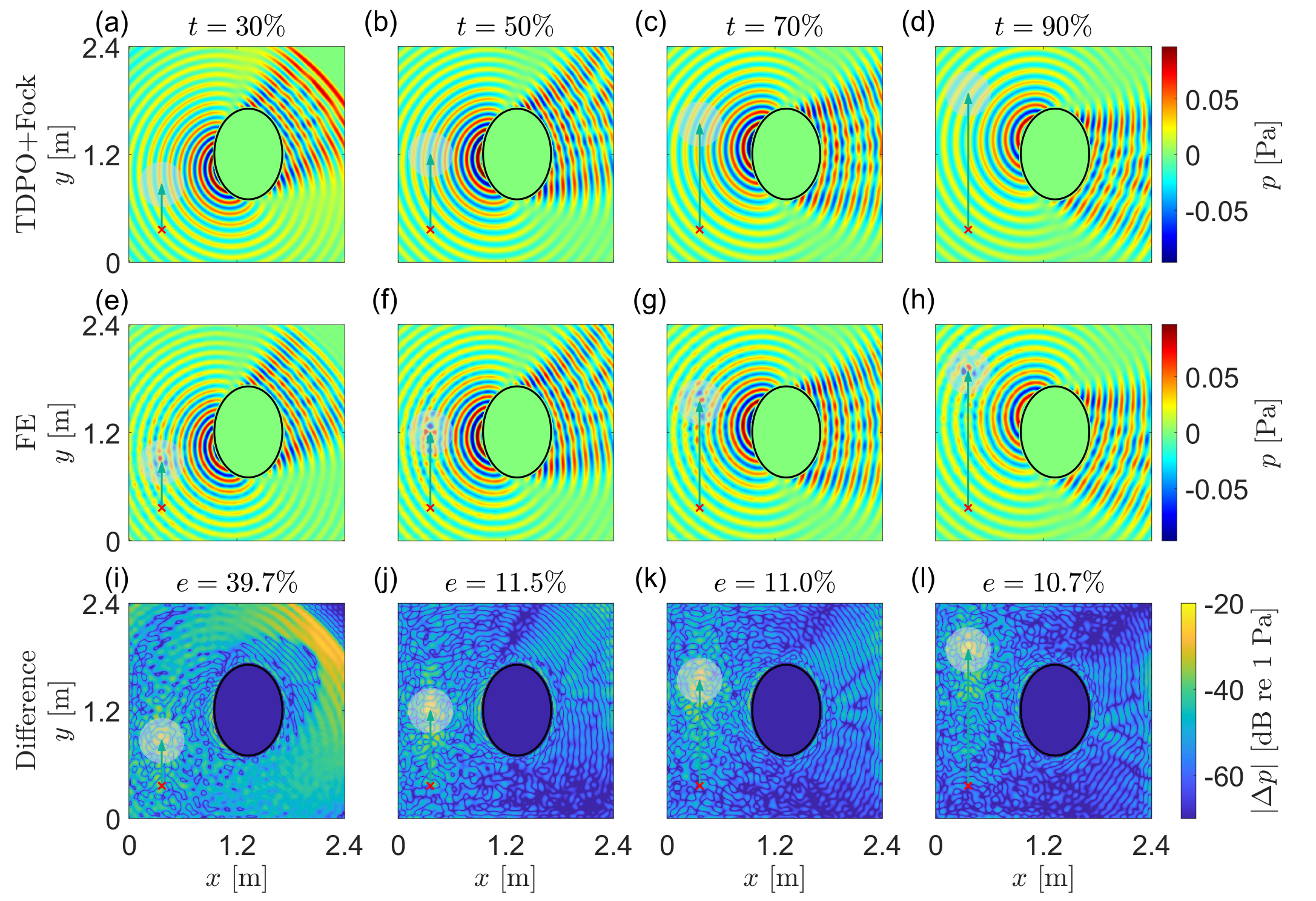}   %
  \caption{\textbf{Moving source and ellipsoid.} Setup of Fig.~\ref{fig:setups}(c). Rows: TDPO+Fock, FE and their difference; columns, markers and $e$ as in Fig.~\ref{fig:sim_sphere}, without the window, with FE as the reference and times given as percentages of $t_{\textrm{end}}=24.5$~ms. The FE field is spatially low-pass filtered to remove its mesh-scale noise, and a disc of radius $2\lambda$ around the source, where the FE point source is inaccurate, is excluded from $e$ and shaded.}
    \label{fig:sim_ellipsoid}
\end{figure*}

Figure~\ref{fig:sim_ellipsoid} compares TDPO with FE. Once the scattered field fills the ROI, the difference stays at about 11\% [Fig.~\ref{fig:sim_ellipsoid}(j)--(l)], close to the FE error for the sphere at $ka=20$ [Fig.~\ref{fig:fd_sphere_diff}(k)], so that it combines the errors of both methods. The two agree best behind the ellipsoid, where the field is formed by the creeping waves, and differ most on the source side. TDPO computes this case in 9~min with 5~GB of memory, whereas the two FE solves, with and without the ellipsoid, take about 96~h with about 300~GB on a similar workstation, or 46~h with up to 168~GB on a GPU, against about 40~s for TDPO on the same GPU.

\section{Discussion and Conclusion}   \label{sec:conclusion}

The TDPO expression for the field scattered by a stationary object illuminated by a moving source was derived, \eqref{eq:p_scat_analytic}--\eqref{eq:chain}. The motion couples the emission, incidence and observation times, and enters the scattered field through the directional Mach number of the source relative to each surface element. Near and beyond the shadow boundary, the PO surface field is supplemented by a Fock correction, extended here to a moving source with a harmonic signal, \eqref{eq:q_fock}--\eqref{eq:p_sh_dot}.

Numerically, the coupled times are handled either backward, by solving for the emission time at each observation time, or forward, by mapping each emission instant to its arrival time and interpolating onto a uniform grid, with no root-finding for any trajectory. The forward approach is used for the scattered field: the incident fields at the surface cells are resampled onto a uniform incidence-time grid, and the field in the ROI is their delay-and-sum, offline or online. 

The validation against the exact solution for a sphere shows that, around the source, where the backscatter dominates, as in sonar, PO alone suffices for acoustically large objects, and that with the Fock correction the error decreases with $ka$, to 3.7\% over the whole ROI and 1.6\% around the source at $ka=30$, below that of FE from $ka=15$ on. This accuracy is retained for a moving source and, near the source, for a broadband signal and a curved trajectory, against an exact solution obtained by superposing stationary-source solutions over the emission instants (Appendix~\ref{app:sphere_exact}). 

For an ellipsoid and a moving source, TDPO agrees with a transient FE simulation to within about 11\%, and takes several minutes and a few gigabytes of memory, against tens of hours and hundreds of gigabytes for FE. TDPO is thus a useful and efficient method for computing the field scattered by an object illuminated by a moving source of arbitrary trajectory and signal, with an accuracy that matches that of FE and, for acoustically large objects, exceeds it.

\begin{acknowledgments}
   \textit{
The authors acknowledge the support of the Pazy Foundation, Award number 348072942, and the support of the Israel Innovation Authority, Award number 84993.}
\end{acknowledgments}

\appendix

\section{Fock correction in the frequency domain} \label{app:fock_fd}

\begin{figure}[!tb]
    \centering
          \includegraphics{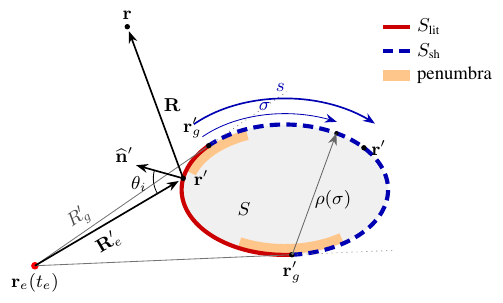}
  \caption{\textbf{Geometry of the Fock correction}, in a cross-section. The grazing points $\mathbf{r}'_g$ form the shadow boundary between $S_{\textrm{lit}}$ and $S_{\textrm{sh}}$. A lit point $\mathbf{r}'$ is illuminated at the angle of incidence $\theta_i$; a shadowed point $\mathbf{r}'$ is reached from $\mathbf{r}'_g$ by a creeping wave along the arc length $s$, with $\sigma$ the arc length to an intermediate point and $\rho(\sigma)$ the local radius of curvature there. The penumbra (orange) is drawn for one wavenumber and is wider at the flatter grazing point.}
    \label{fig:fock_geometry}
\end{figure}

For a smooth convex rigid object, Fock's theory \cite{Fock1965,Pathak1980} replaces the PO surface field \eqref{eq:PO_def} by
\begin{equation} \label{eq:fock_surface}
\widetilde{p}_{\textrm{tot}}(\mathbf{r}')\approx
\begin{cases}
F_{\textrm{lit}}\big(\xi_{\textrm{lit}}(\mathbf{r}')\big)\,\widetilde{p}_{\textrm{inc}}(\mathbf{r}'), & \mathbf{r}'\in S_{\textrm{lit}},\\[2pt]
\mathcal{D}(\mathbf{r}')\,F_{\textrm{sh}}\big(\xi_{\textrm{sh}}(\mathbf{r}')\big)\,\widetilde{p}_{\textrm{inc}}(\mathbf{r}'_g)\,e^{-\mathrm{j}ks(\mathbf{r}')}, & \mathbf{r}'\in S_{\textrm{sh}},
\end{cases}
\end{equation}
where $S_{\textrm{sh}}$ is the shadowed part of $S$ (Fig.~\ref{fig:fock_geometry}), and $s$ is the arc length along the surface geodesic that continues the grazing ray to $\mathbf{r}'$. The factor $\mathcal{D}$, equal to unity at $\mathbf{r}'_g$, accounts for the spreading of the creeping waves and for the change of curvature along the path \cite{Pathak1980}. On a sphere, $\mathcal{D}=(\sin\alpha_g/\sin\alpha)^{1/2}$, with $\alpha$ the angle at the center between $\mathbf{r}'$ and the source direction and $\alpha_g$ its value on the boundary.
The transition takes place over the penumbra, which narrows as the frequency increases and extends about $\rho\,(2/k\rho)^{1/3}$ on either side of the boundary, with $\rho$ the radius of curvature along the direction of incidence. The distance from the boundary is measured in penumbra widths by
\begin{equation} \label{eq:xi_fock}
\begin{aligned}
\xi_{\textrm{lit}}(\mathbf{r}')&=-\left(\frac{k\rho(\mathbf{r}')}{2}\right)^{1/3}\cos\theta_i(\mathbf{r}'),\\
\xi_{\textrm{sh}}(\mathbf{r}')&=\int_0^{s(\mathbf{r}')}\left(\frac{k}{2\rho^2(\sigma)}\right)^{1/3}\mathrm{d}\sigma,
\end{aligned}
\end{equation}
which is negative on $S_{\textrm{lit}}$ and positive on $S_{\textrm{sh}}$. Here, $\cos\theta_i=-\widehat{\mathbf{R}}'_e\cdot\widehat{\mathbf{n}}'$, $\sigma$ is the arc length along the geodesic, and $\rho(\sigma)$ is the radius of curvature along it. On a sphere of radius $a$, $\xi_{\textrm{sh}}=(ka/2)^{1/3}s/a$. The penumbra is the band $|\xi|\lesssim1$. Fock's factors for a rigid surface are
\begin{equation} \label{eq:fock_factors}
F_{\textrm{sh}}(\xi)=\frac{1}{\sqrt{\pi}}\int_{\Gamma}\frac{e^{-\mathrm{j}\xi\zeta}}{w'(\zeta)}\,\mathrm{d}\zeta,\qquad F_{\textrm{lit}}(\xi)=e^{-\mathrm{j}\xi^3/3}F_{\textrm{sh}}(\xi),
\end{equation}
with the Fock--Airy function $w(\zeta)=\sqrt{\pi}\,[\mathrm{Bi}(\zeta)-\mathrm{j}\,\mathrm{Ai}(\zeta)]$, $\mathrm{Ai}$ and $\mathrm{Bi}$ the Airy functions of the first and second kind, $w'=\mathrm{d}w/\mathrm{d}\zeta$, and $\Gamma$ running from infinity along $\arg\zeta=-5\pi/6$ to the origin and out along the positive real axis \cite{Pathak1980}. Both factors equal about 1.4 on the boundary; deep in the lit region, $F_{\textrm{lit}}\approx2-\mathrm{j}/(2\xi^3)$ tends to the PO value, and in the shadow $F_{\textrm{sh}}$ is a sum of exponentially decaying creeping modes.
Since the factors depend on the surface point only through $\xi(\mathbf{r}')$, substituting \eqref{eq:fock_surface} into \eqref{eq:p_scat} multiplies the integrand of \eqref{eq:p_scat_final} by $F_{\textrm{lit}}/2$ on $S_{\textrm{lit}}$, and adds an integral over $S_{\textrm{sh}}$ of the same form, weighted by $\mathcal{D}F_{\textrm{sh}}/2$, with $R'_e$ replaced by $R'_g$ in the amplitude and by $R'_g+s$ in the phase.

\section{Finite element simulations} \label{app:fe}

FE simulations are carried out with COMSOL Multiphysics~6.4 for the stationary source and sphere of Sec.~\ref{sec:sim_fd_sphere} and for the moving source and ellipsoid of Sec.~\ref{sec:sim_ellipsoid}. In both, the domain, a box of $2.4\times2.4\times1.4$~m that contains the ROI and the object and is surrounded by a perfectly matched layer one wavelength thick, is halved by the symmetry plane $z=0$, with the monopole driven at half its strength. 

The fields of Sec.~\ref{sec:sim_fd_sphere} are computed in the frequency domain. For each $ka$, the incident and total fields are computed without and with the sound-hard sphere, and their difference on $z=0$ is interpolated onto the ROI grid. The number of degrees of freedom grows from $2.5\times10^{6}$ at $ka=10$ to $4.8\times10^{7}$ at $ka=30$. Since the element size is fixed in wavelengths while the ROI spans more wavelengths as $ka$ increases, the numerical dispersion accumulates, and the FE error grows with $ka$ [Fig.~\ref{fig:fd_sphere_diff}(i)--(l)].

For the moving source of Sec.~\ref{sec:sim_ellipsoid}, FE requires a transient simulation with the source position updated at every time step. The monopole moves over a fixed mesh of elements of $\lambda/5$, i.e., $2.4\times10^{7}$ degrees of freedom with the ellipsoid, and the generalized-$\alpha$ time step is $T/50$. As for the stationary source, the scattered field is the difference of two solves, without and with the ellipsoid. Over the 24.5-ms record, each solve takes tens of hours and hundreds of gigabytes of memory, on a CPU as well as on a GPU.

\section{Exact solution for a sphere} \label{app:sphere_exact}

For a stationary monopole at $\mathbf{r}_e$ radiating a harmonic signal, the field scattered by a rigid sphere of radius $a$ centered at $\mathbf{r}_c$ is $\widetilde{p}_{\textrm{scat}}(\mathbf{r};\omega)=G_{\textrm{s}}(\mathbf{r},\mathbf{r}_e;\omega)\,\widetilde{q}_e$. The scattered Green's function $G_{\textrm{s}}$ is obtained by expanding the incident field \eqref{eq:p_inc} in spherical waves about $\mathbf{r}_c$ and imposing a zero normal derivative of the total field on the sphere, which gives the partial-wave series~\cite{bowman1970scattering}
\begin{equation} \label{eq:G_sphere}
    G_{\textrm{s}}=\frac{\mathrm{j}k}{4\pi}\sum_{\ell=0}^{\infty}(2\ell+1)\frac{j_\ell'(ka)}{h_\ell'(ka)}h_\ell(kd_e)h_\ell(kd)P_\ell(\cos\psi).
\end{equation}
Here, $d=|\mathbf{r}-\mathbf{r}_c|$ and $d_e=|\mathbf{r}_e-\mathbf{r}_c|$ are the distances from the sphere center, and $\psi$ is the angle between $\mathbf{r}-\mathbf{r}_c$ and $\mathbf{r}_e-\mathbf{r}_c$. The functions $j_\ell$ and $h_\ell=j_\ell-\mathrm{j}y_\ell$ are the spherical Bessel and Hankel functions, the latter outgoing for the time dependence $e^{\mathrm{j}\omega t}$, and $P_\ell$ is the Legendre polynomial; the prime denotes the derivative. The series holds everywhere outside the sphere. Since $j_\ell'(ka)/h_\ell'(ka)$ decays faster than exponentially once $\ell$ exceeds $ka$, it is truncated well beyond $\ell\approx ka$, where the terms are negligible.

For a moving source with any signal $q_e(t_e)$, the sphere is fixed and the problem is linear, so that the scattered field is the superposition, over the emission instants $t_e$, of the fields scattered by a stationary monopole at $\mathbf{r}_e(t_e)$,
\begin{multline} \label{eq:p_scat_exact}
    p_{\textrm{scat}}(\mathbf{r},t)=\frac{1}{2\pi}\int_{-\infty}^{\infty}\!\int_{-\infty}^{\infty}q_e(t_e)\\
    \times G_{\textrm{s}}\!\left(\mathbf{r},\mathbf{r}_e(t_e);\omega\right)e^{\mathrm{j}\omega(t-t_e)}\,\mathrm{d}t_e\,\mathrm{d}\omega.
\end{multline}
Replacing $G_{\textrm{s}}$ by $G(\mathbf{R}_e;\omega)$ in \eqref{eq:p_scat_exact} recovers the incident field \eqref{eq:p_inc_time}, including the Doppler amplitude factor, and for a stationary harmonic source, \eqref{eq:p_scat_exact} reduces to $\widetilde{p}_{\textrm{scat}}=G_{\textrm{s}}\widetilde{q}_e$.

Since $q_e(t_e)=0$ before the emission starts, and the field scattered by the sphere from an impulse emitted at $t_e$ arrives at $\mathbf{r}$ after a propagation delay, the field at time $t$ is formed by emissions preceding $t$. In TDPO, the free-space impulse response, a delta function, reduces this superposition to a single emission instant per surface point, \eqref{eq:t_e_t'}; the scattered impulse response of the sphere is spread in time, and \eqref{eq:p_scat_exact} is evaluated offline, over the entire record.

\bibliography{TDPO}

\end{document}